\documentclass[sigconf]{acmart}
\usepackage{listings}
\usepackage[htt]{hyphenat}
\usepackage{changepage} 
\usepackage{courier}
\usepackage[T1]{fontenc}
\AtBeginDocument{%
  \providecommand\BibTeX{{%
    \normalfont B\kern-0.5em{\scshape i\kern-0.25em b}\kern-0.8em\TeX}}}

\usepackage{csquotes}
\setcopyright{acmlicensed}
\copyrightyear{2018}
\acmYear{2018}
\acmDOI{XXXXXXX.XXXXXXX}

\acmConference[DEEM@Sigmod 2024]{Pre-print}{June 09--14,
  2024}{Santiago, Chile}
\acmISBN{978-1-4503-XXXX-X/18/06}

\begin{document}


\title{Reproducible data science over data lakes: replayable data pipelines with \texttt{Bauplan} and \texttt{Nessie}.}

\author{Jacopo Tagliabue}
\email{jacopo.tagliabue@bauplanlabs.com}
\orcid{0000-0001-8634-6122}
\authornote{This is a pre-print, non-final version of the paper accepted at DEEM SIGMOD 2024. For the final archival version, please check the official proceedings after the conference.}
\affiliation{%
  \institution{Bauplan, NYU Tandon}
  \city{New York}
  \country{USA}
}

\author{Ciro Greco}
\email{ciro.greco@bauplanlabs.com}
\affiliation{%
  \institution{Bauplan}
  \city{New York}
  \country{USA}
}

\renewcommand{\shortauthors}{Tagliabue, Greco}

\begin{abstract}
    As the Lakehouse architecture becomes more widespread, ensuring the reproducibility of data workloads over data lakes emerges as a crucial concern for data engineers. However, achieving reproducibility remains challenging. The size of data pipelines contributes to slow testing and iterations, while the intertwining of business logic and data management complicates debugging and increases error susceptibility. In this paper, we highlight recent advancements made at Bauplan in addressing this challenge. We introduce a system designed to decouple compute from data management, by leveraging a cloud runtime alongside Nessie, an open-source catalog with Git semantics. Demonstrating the system's capabilities, we showcase its ability to offer time-travel and branching semantics on top of object storage, and offer full pipeline reproducibility with a few CLI commands.
\end{abstract}



\maketitle

\section{Introduction}

\begin{displayquote}
``No man ever steps in the same river twice, for it's not the same river and he's not the same man'' -- \textit{Heraclitus}
\end{displayquote}

Reproducibility is always mentioned as a major obstacle in debugging data science projects and in moving them from development to production \cite{Martinez2021ASS,https://doi.org/10.48550/arxiv.2209.09125}. 

The conventional engineering approach, which is based on replicating computer behavior by repeatedly inputting the same data into the same code, reveal critical limitations when confronted with modern data workloads. As shown in Table \ref{tab:checklist}, reproducing a data pipeline needs versioning and portability of extensive inputs, modular code, runtime compatibility with various packages, and hardware flexibility. While existing tools may function adequately in isolation, enabling time-travel capabilities across all these components to reproduce data pipelines demands substantial engineering proficiency, setup and context-switching.

In \textit{this} short paper we describe the recent progress we made at \textit{Bauplan} in attaining reproducibility through a unified framework for data pipelines over \textit{data lakes}. In particular, we demonstrate how a system based on declarative pipelines can decouple the business logic from runtime and data management, and address the challenges illustrated in 
Table \ref{tab:checklist}. We summarize our contributions as follows:

\begin{enumerate}
    \item we outline abstractions that allow data pipelines to be implemented in multiple languages and artifacts to be represented transparently across the hierarchy of persistence (in-memory tables, parquet files, data lake tables and data branches);

    \item we explain the architecture and the ergonomics of our CLI, which allows practitioners to write pipelines in their local IDE and run them in directly the cloud through Bauplan's FaaS runtime;
    
    \item we describe the open-source Nessie data catalog and show how Git-like semantics can be applied to datasets over data lake.
\end{enumerate}

\begin{table}
  \caption{Reproducibility checklist}
  \label{tab:checklist}
  \begin{tabular}{cccc}
    \toprule
    Component & Tools & Example\\
    \midrule
    \textit{input data} & S3, Iceberg & 100M row dataframe\\
    \midrule
    \textit{code} & Git & 10+ SQL / Python functions\\
    \midrule
    \textit{runtime} & pip, Docker & \texttt{scikit==1.3.0}\\
    \midrule
    \textit{hardware} & local machine, cloud & EC2\\
  \bottomrule
\end{tabular}
\end{table}

\section{Data pipelines as functional DAGS}
\label{sec:abstractions}

\begin{figure*}
    \centering
    \includegraphics[width=\linewidth]{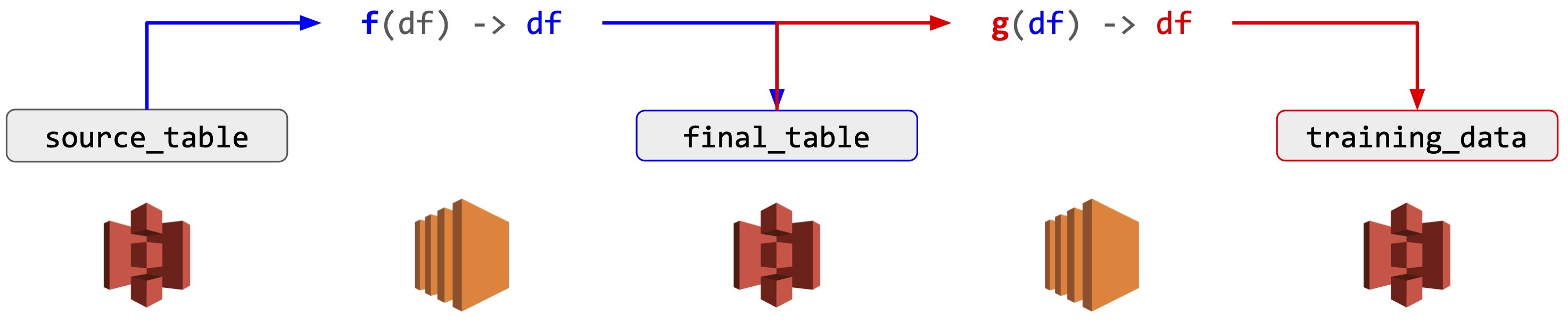}
    \caption{Interleaving of compute (EC2) and storage (S3) in a data pipeline, modelled as a Directed Acyclic Graph (DAG): the responsibility of the data scientist is to write functions that transform the original data artifact (the source data) in intermediate and then final dataframes for downstream consumption (e.g. run a ML model).}    
    \label{fig:dag}
\end{figure*}

Data pipelines are the bread and butter of any data processing, serving ETL, analytics, reports and model building. Let's discuss a typical use case, for illustrative purposes:
\\
\\
\textbf{Example use case \#1}: Richard is a data scientist at \textit{ACME Inc.}, a financial company interested in detecting fraudulent transactions on its platform. Richard is tasked with developing pipeline \textit{P}  (Fig.~\ref{fig:dag}) to transform the raw transaction logs into a clean tabular structure suitable for further analysis and ML workloads.
\\
\\
Fig.~\ref{fig:dag} showcases two important concepts: 
    \begin{enumerate}
        \item The \textit{separation between storage and compute}, as encouraged by data lake architectures (e.g. the implementation of \textit{P} would look different in a traditional database like PostgreSQL, or a cloud warehouse like Snowflake). This architecture is the focus of the current system, and it is prevalent in most mid-to-large enterprises (its benefits that been discussed at length before \cite{Zaharia2021LakehouseAN,mazumdar2023data}).
        \item The \textit{functional nature of \textit{P}}: transformation functions (e.g. \textit{g} in Fig.~\ref{fig:dag}) are not required to know anything about how artifacts are created or persisted by previous transformation steps. As long as the process computing \texttt{g} receives the ``right input'', \texttt{g} and \texttt{f} can be run in completely different environments, with different languages.  
    \end{enumerate}

These two points imply that moving data from compute to storage and vice-versa can be decoupled entirely from the transformation logic. Ideally, Richard can focus on writing \texttt{g} and \texttt{f}, while the underlying infrastructure automatically handles compression, serialization, movement and persistence. If we think of the ``tabular structures'' as \textit{dataframes}\cite{10.14778/3407790.3407807}, \textit{P} can be modelled as DAGs in which nodes are dataframes, and edges are transformation functions (from dataframes to dataframes): running \textit{P} successfully is semantically equivalent to the composition of $training\_data=g(f(source\_table))$); in turn, running \textit{f} or \textit{g} successfully depends only on their input dataframes having the right semantics, as encoded in the compatibility of their respective schema. Richard can reason about \textit{P} entirely at the schema level, moving between languages as he sees fit and leaving the data representation hierarchy as an implementation detail (Fig.~\ref{fig:hierarchy}).

The listings below provide a multi-language \textit{P} implementation with \textit{Bauplan} syntax: the simplicity of the chosen abstractions removes the burden of translating from SQL tables to Pandas, serializing / deserializing dataframes, reading / writing from / to S3 efficiently:

\begin{itemize}
    \item \texttt{final\_table} (the intermediate dataframe in \textit{P}) is produced through a SQL file with the same name\footnote{Note here the adoption of the popular \textit{dbt} naming convention: \url{https://github.com/dbt-labs/dbt-core}.}, querying \textit{FROM} the raw data in the source table, thereby implicitly declaring its parent;
    \item \texttt{training\_data} (the final dataframe) is produced through a Python function with the same name, accepting as input the intermediate table, thereby implicitly declaring its parent.
\end{itemize}

\begin{lstlisting}[columns=fullflexible,language=SQL,caption=final\_table.sql]
SELECT c1, c2, c3
FROM source_table -- reference to its parent DAG node
WHERE transactionDate >= DATEADD(day,-7, GETDATE())
\end{lstlisting}

\begin{lstlisting}[columns=fullflexible,language=Python,caption=training\_data.py]
@bauplan.model()
@bauplan.python('3.11', pip={'scikit-learn': '1.3.0'})
def training_data(
    # reference to its parent DAG node
    data=bauplan.Model('final_table') 
):
   # transformation logic DAG here
   my_final_df = data.do_something()
   return my_final_df
\end{lstlisting}

In this perspective, assuming we can count on a cloud runtime that supports the right dependencies (e.g. the function requires \texttt{scikit-learn}), we can deterministically reproduce any past instance of $training\_data$ simply with the code above and an immutable reference to the source data in S3. In other words, to provide a comprehensive treatment of the reproducibility checklist in Table \ref{tab:checklist}, we would need a platform that provides \textit{i)} data and code versioning, and \textit{ii)} a cloud runtime.

In the ensuing sections, we outline our solution for the two missing pieces of the puzzle: first, how to go from dataframes to files and vice versa, because while Richard thinks in terms of the former, data versioning on S3 happens in terms of the latter (Section \ref{sec:catalog}); second, how to run Richard's code directly in the cloud, to avoid runtime and hardware discrepancies when reproducing pipeline runs (Section \ref{sec:bauplan}). 

\section{FROM DATAFRAMES TO A DATA CATALOG}
\label{sec:catalog}

\begin{figure}
  \centering
  \includegraphics[width=0.8\linewidth]{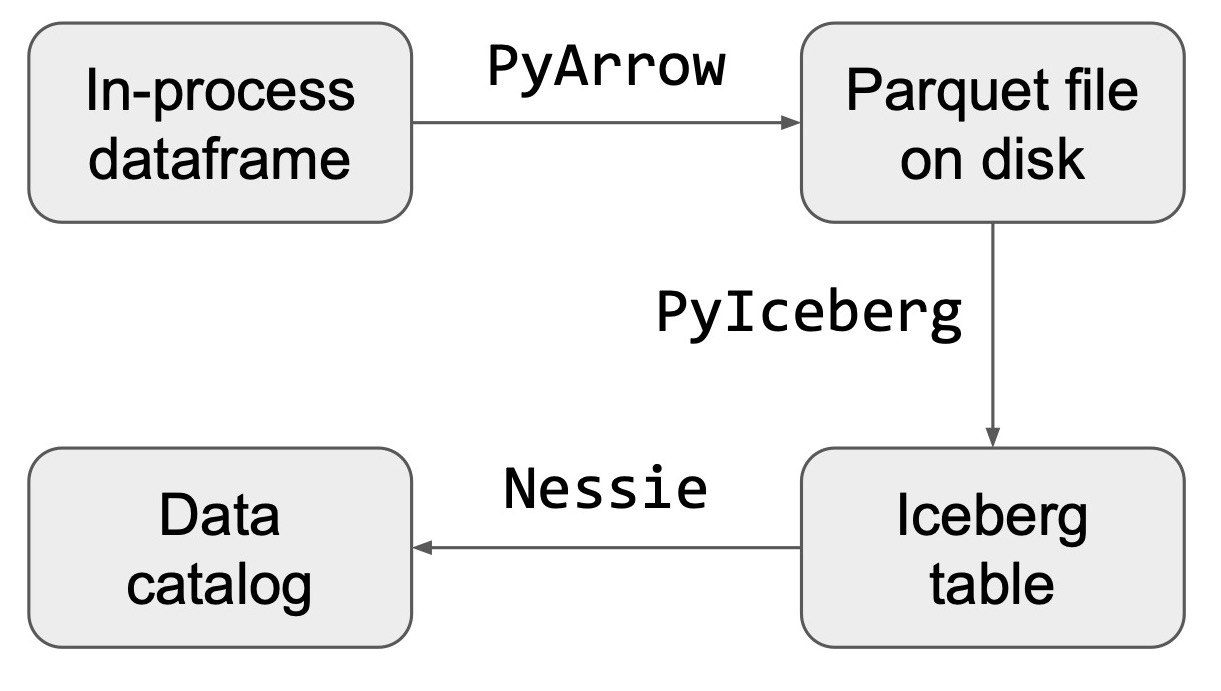}
  \caption{Hierarchy of data representation: while data scientists interact only with in-process dataframes, the system persists the information preserving the overall semantics through different (reversible) layers of abstraction -- physical files, collection of files into tables, collection of tables.}
  \label{fig:hierarchy}
\end{figure}

As the architecture fully embraces the separation of storage and compute, the only way to persist states is through object storage -- according to our abstractions, the relevant states are the dataframes produced by transformation functions, and storage is provided by S3 (or equivalent service). Fig. \ref{fig:hierarchy} illustrates the write path that starts from the the transformation in \textbf{training\_data.py}  -- a Python dataframe that only lives inside the runtime memory --, and ends with a table in S3 -- an \textit{Iceberg} table that can be now queried by downstream pipelines, or by any compatible query engine (\textit{duckdb}, \textit{Snowflake}, \textit{Dremio}, etc.):

\begin{enumerate}
    \item \textbf{from Arrow to Parquet}: standard open source libraries\footnote{\url{https://arrow.apache.org/docs/python/index.html}} perform the conversion from the in-memory representation (Arrow) to the compressed one (Parquet);
    \item \textbf{from Parquet to Iceberg}: when a dataframe is converted to multiple files in S3, metadata are required to preserve a dataframe semantics for downstream systems. We turned to the popular open format \textit{Iceberg}\footnote{\url{https://iceberg.apache.org/}}, which defines metadata to contain the \textit{schema} (common to all files) and pointers to row groups as physically stored in S3. This level of indirection enables transaction-like behavior over the data lake: users can reason with high-level abstractions such as schema evolution and table snapshots, instead of tracking file changes. Inserts and updates produce a unique commit, which can be referenced as an immutable state of the table;
    \item \textbf{from Iceberg to Nessie}: multiple \textit{Iceberg} tables are managed by a further abstraction, a data catalog -- we picked Nessie\footnote{\url{https://projectnessie.org/}} as our open source catalog, for its support for multi-table transactions (crucial for data pipelines) and data branching. Branching and merging for dataframes are analogous to Git operations, and allow Richard to launch runs from the same data \texttt{source\_table} has, but sandboxing his transformations to avoid pushing to downstream services bad data while he's developing. While branching is not necessary for reproducibility, it's an essential component (with reproducibility) for safe debugging (Section \ref{sec:repro}).
\end{enumerate}

On the read path the system performs the same conversion in reverse order: given a branch selected by Richard (Section \ref{sec:repro}), it identifies the relevant table commits, retrieves the files and convert Parquet to Arrow to set the input for the next function.

\section{BAUPLAN ARCHITECTURE}
\label{sec:bauplan}

If the code in Section~\ref{sec:abstractions} captures the first person perspective of \textit{writing} pipelines, we now have to show the first person perspective of \textit{running} them. From the point of view of Richard, \textit{Bauplan} is a pip-installable package, which gives access to lakehouse capabilities\cite{10.14778/3603581.3603604,Zaharia2021LakehouseAN} through simple CLI-based interactions: after writing the code for \textit{P}, Richard can execute it in the cloud with a simple terminal command: \texttt{bauplan run}.

\begin{figure}
  \centering
  \includegraphics[width=\linewidth]{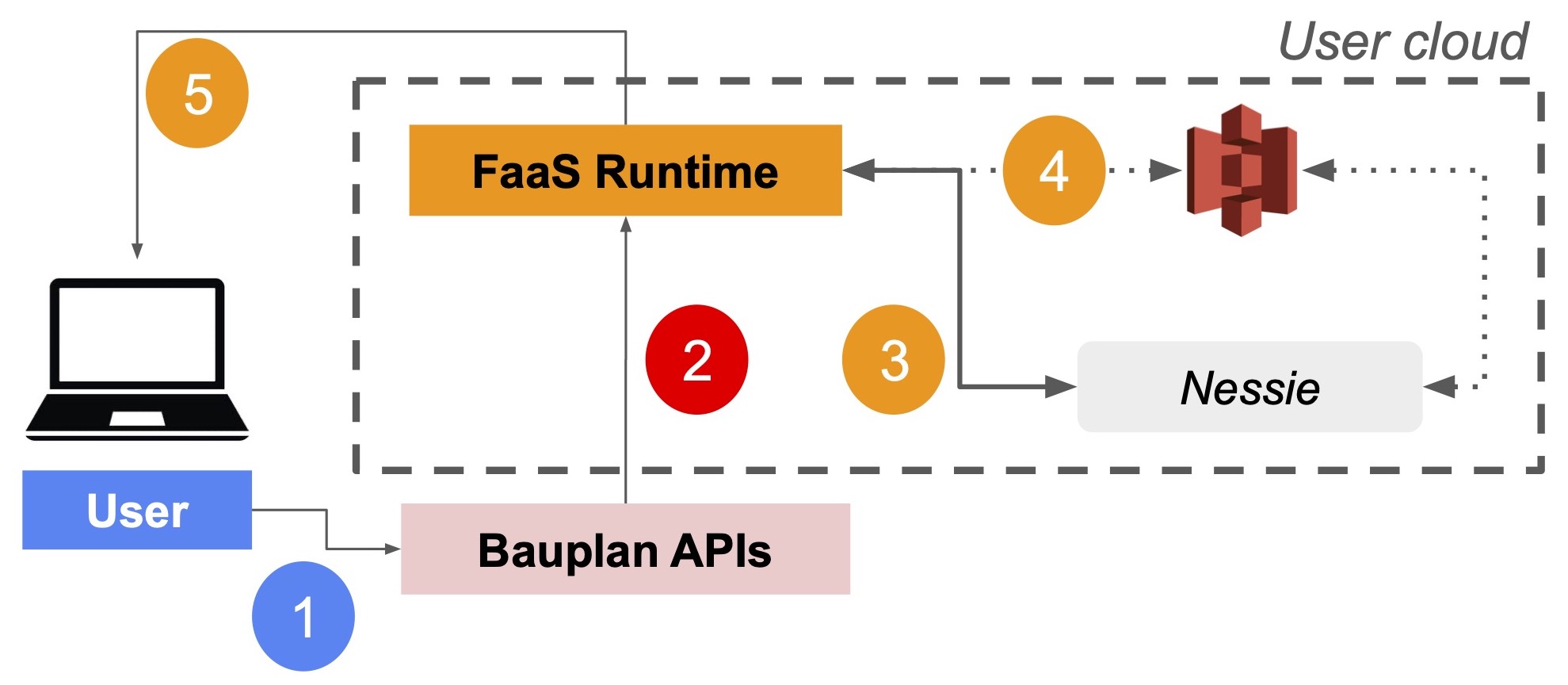}
  \caption{Data flow for a pipeline run: 1) user issues a query to the middleware, which 2) sends a plan to the runtime; 3) the runtime asks Nessie for the parquet files backing the plan and 4) retrieves them from S3. Finally, the pipeline is run and 5) results are sent back to the client.}
  \label{fig:system}
\end{figure}

Fig.\ref{fig:system} describes the system perspective of a run: the code gets sent to the API, which parses it and outputs a plan for the execution in Richard's cloud\footnote{In a lakehouse, data can be processed directly in the target cloud, with no additional costs or privacy concerns.}; the runtime communicates with \textit{Nessie} to retrieve the appropriate data from object storage, and then executes the plan by converting files to dataframes for the pipeline nodes.\footnote{Details on the custom FaaS runtime powering \textit{Bauplan} are beyond the scope of \textit{this} paper, but the interested reader may check \cite{Tagliabue2023BuildingAS,Tagliabue2024}.} Two observations are worth mentioning:

\begin{itemize}
    \item \textbf{pipeline abstractions are naturally converted to ``Function as a Service'' (FaaS) execution}: if pipelines are DAGs of transformations, FaaS semantics makes for a great developer interface; furthermore, FaaS emphasizes a clear separation of concerns between users (Richard writing \textit{f} and \textit{g}) and the system (converting dataframes from / to memory, scheduling function execution efficiently etc.);
    \item \textbf{runs are immutable}: every \texttt{bauplan run} returns a \texttt{run-id}, which uniquely identifies the combination of the code \textit{and} the input data (the data commit mentioned above).
\end{itemize}

Crucially, since \textit{Bauplan} sits at the intersection of code, runtime and data access, it can provide \textit{through a single abstraction} the time-travel capabilities of what typically happens in separate systems (Table \ref{tab:checklist}). We will now show how these building blocks can solve efficiently a typical debugging use case.

\section{REPRODUCING PIPELINES}
\label{sec:repro}

\begin{figure}
  \centering
  \includegraphics[width=\linewidth]{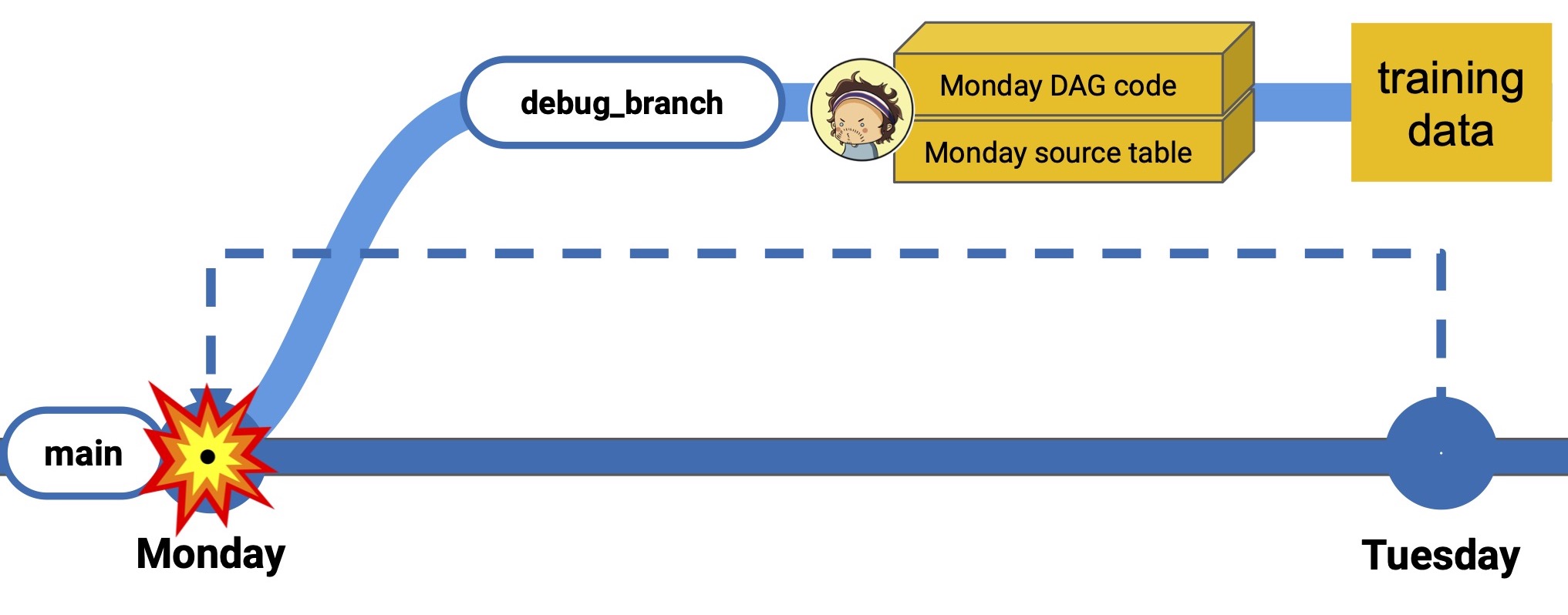}
  \caption{Debugging on \textit{Bauplan}: when Richard reproduces Monday's run, the system 1) travels back in time at Monday's source data \textit{and} pipeline code, 2) creates a debug branch for his experiments, and 3) materializes the target artifacts inside the branch.}
  \label{fig:branch}
\end{figure}

Let's consider this common debugging use case:
\\
\\
\textbf{Example use case \#2}: \textit{P} now runs in production every night\footnote{Scheduling details are not important for the example: you can imagine executing a \textit{Bauplan} run from a cron job or through more complex orchestration.} -- when it ran last night, it unexpectedly produced an empty \texttt{training\_data} table. Richard is tasked to identify and fix the bug.
\\
\\
The use case brings two connected but distinct challenges ((Fig.~\ref{fig:branch}). First, there is \textit{reproducibility}: to reproduce the faulty run, Richard needs to run the same code over the same source data as last night; then, there is \textit{materialization}: when debugging past runs, Richard should have a temporary version of \texttt{training\_data}, so that his debugging attempts won't interact with production artifacts that the rest of the company is using. Since \textit{Bauplan} provides immutable reference to code and input data for every run, the solution to both our challenges is a few CLI commands away:

\begin{lstlisting}[columns=fullflexible,language=bash,caption=Reproducing a pipeline (CLI)]
bauplan checkout richard.debug_branch
bauplan run --id=1441804 
bauplan query "SELECT COUNT(*) FROM training_data"
\end{lstlisting}

Even if the reader never encountered the commands before, its semantics should be obvious: given the id of last night production run ($1441804$), Richard can: 1) create a target branch separate from production to host dataframes while debugging; 2) re-run last night pipeline starting from the same input and re-using the same code (ensuring reproducibility); 3) query a dataframe in his branch, to reproduce the bug first and then verify how the final table changes as he fixes the code\footnote{In other words, \texttt{COUNT} should be zero at first when the exact production run gets replayed, and then it changes as Richard starts fixing the underlying cause during iterative debugging.}. Some points are worth highlighting:

\begin{enumerate}
    \item \textbf{CLI is all you need}: Richard does not need to know / setup / provision a data catalog service, nor learn its API, download a client etc. a simple Git-like command is enough for him to operate the system proficiently and achieve the intended goal;
    \item \textbf{built-in namespacing}: we follow a \texttt{user.branch} convention, so that users can only write in their branches, but everybody can read any branch;
    \item \textbf{interoperability with query engines}: artifacts can be queried within the platform itself through SQL with no additional setup, or they can be read by any Iceberg-compatible engine;
    \item \textbf{efficient data re-use}: when branching occurs (Fig.~\ref{fig:branch}), the original source table at the start of the DAG is not copied: \textit{Nessie} builds the debug branch through copy-on-write semantics over the lake, avoiding slow and costly copies;
    \item \textbf{extensibility to CI/CD}: the same building blocks can be used to enforce a Write-Audit-Publish pattern during normal development, or even during scheduled execution: a common pattern among \textit{Bauplan} users is to run Python tests over dataframes for data quality\footnote{These are typically called \textit{expectations}, and they are functions from dataframes to booleans.}; branching, testing and merging through a command line API allow a CI/CD similar to software builds.
\end{enumerate}

Taken all together, \textit{Bauplan} APIs provide a unified, multi-language abstraction over the four reproducibility components in Table \ref{tab:checklist}: \textit{code} and \textit{input data} are versioned at each \texttt{bauplan run}, leveraging Nessie and Bauplan own APIs; \textit{runtime}'s concerns are expressed directly in code (as required Python packages), \textit{hardware} is stable across runs, as local execution is avoided altogether thanks to the FaaS cloud engine.

\section{Related work}

DAG-based modelling of data pipelines is common in orchestrators (e.g. Airflow\footnote{\url{https://airflow.apache.org/}}): unlike Bauplan however, orchestrators do not provide built-in runtimes nor direct access to dataframe semantics, leaving users to roll their own reproducibility recipe by stitching together several tools.

\textit{Dvc} is a popular ``Data Version Control'' system, which primarily operates with file semantics: as a consequence, the system is mostly used for local files and single-file datasets,\footnote{\url{https://dvc.org/doc/user-guide/data-management/importing-external-data}} as opposed to the thousands of files composing Iceberg tables on a lake. The lack of dataframe semantics is reflected in basic usage of S3 and lack of interoperability with query engines. 

\textit{Metaflow} provides S3-based immutable runs \cite{Tagliabue2023ReasonableSM}, and it is (to the best of our knowledge) the only other system versioning code and data artifacts; its abstractions are however more generic, and data management is entirely left to the users: since any variable is a blob, dataframes are not interoperable with SQL engines nor they are addressable directly with table semantics.

\section{CONCLUSION AND FUTURE WORK}
Reproducing pipelines over a data lake is a common concern for modern enterprises: the absence of table semantics and the complexity of inter-operating code, runtime and storage, poses a formidable challenge for practitioners. We presented \textit{Bauplan} and \textit{Nessie} as an alternative to fragmented tooling: by combining in a simple API a multi-language cloud runtime and data branching, we obtained full reproducibility with just a few CLI commands.


\bibliographystyle{ACM-Reference-Format}
\bibliography{sample-base}

\end{document}